\documentclass[11pt]{article}
\usepackage{moriond,epsfig}
\usepackage{amssymb}

\def\ra{\rightarrow}

\def\be{\begin{equation}}
\def\ee{\end{equation}}
\def\bea{\begin{eqnarray}}
\def\eea{\end{eqnarray}}

\begin{document}
\vspace*{4cm}
\title{Observability of MSSM Higgs bosons decaying to sparticles at the LHC}

\author{ Filip Moortgat }

\address{Department of Physics, University of Antwerpen, \\
Universiteitsplein 1, B-2610 Wilrijk, Belgium}

\maketitle\abstracts{
The possibility is discussed to observe the sparticle decay modes of the heavy 
MSSM Higgs bosons at the Large Hadron Collider.
We focus on the heavy neutral Higgses $H^0$ and $A^0$, and argue that their 
decay into neutralinos may access an interesting region in MSSM parameter 
space up to $m_A$ = 450 GeV for low and intermediate values of $\tan \beta$. 
If neutralinos and sleptons are light enough, the $H^0, A^0 
\ra \chi^0_2 \chi^0_2 \ra 4l^{\pm} + X$ channel can complement 
the reach of the SM channels.}

\section{Introduction}

In order to explain the mechanism of electroweak symmetry breaking in the 
Minimal Supersymmetric Standard Model (MSSM), the existence of two Higgs 
doublets is assumed, leading to five physical Higgs bosons: a light CP-even ($h^0$), 
a heavy CP-even ($H^0$), a heavy CP-odd ($A^0$) and two charged Higgs bosons ($H^{\pm}$).
A discovery of one of these heavy Higgses would therefore be a major breakthrough in
verifying the supersymmetric nature of the underlying theory.
At the Large Hadron Collider (LHC), the most promising channels to discover the
heavy Higgs bosons seem to be the $H^0, A^0 \ra \tau \tau$ or the $H^+ \ra \tau \nu$ channel. 
These channels \cite{kin} were studied extensively in the CMS and ATLAS Collaborations and can lead 
to discoveries in the large and intermediate $\tan\beta$ region of the MSSM parameter space. 
In the studies of these channels, it was assumed that sparticles are heavy ($\sim$ 1 TeV), and 
do not participate in the decay process. However, if neutralinos and/or charginos would be light,
the branching ratios into these sparticles will be sizable. Therefore we investigate whether 
there is a way to observe such $H^0, A^0 \ra \chi^+ \chi^- , \chi^0 \chi^0$
decays.
We will show that in the low and intermediate $\tan\beta$ region of the MSSM
parameter space, this might indeed be the case
for 250 $\lesssim$ $m_A$ $\lesssim$ 450 GeV, depending on the values of $M_2$, $\mu$ and 
$m_{\tilde{l}}$.

\section{Neutralino/chargino decay modes at the LHC}
\subsection{Branching ratios}
\begin{figure}
\begin{minipage}[l]{65mm}
\begin{center}
\epsfig{file=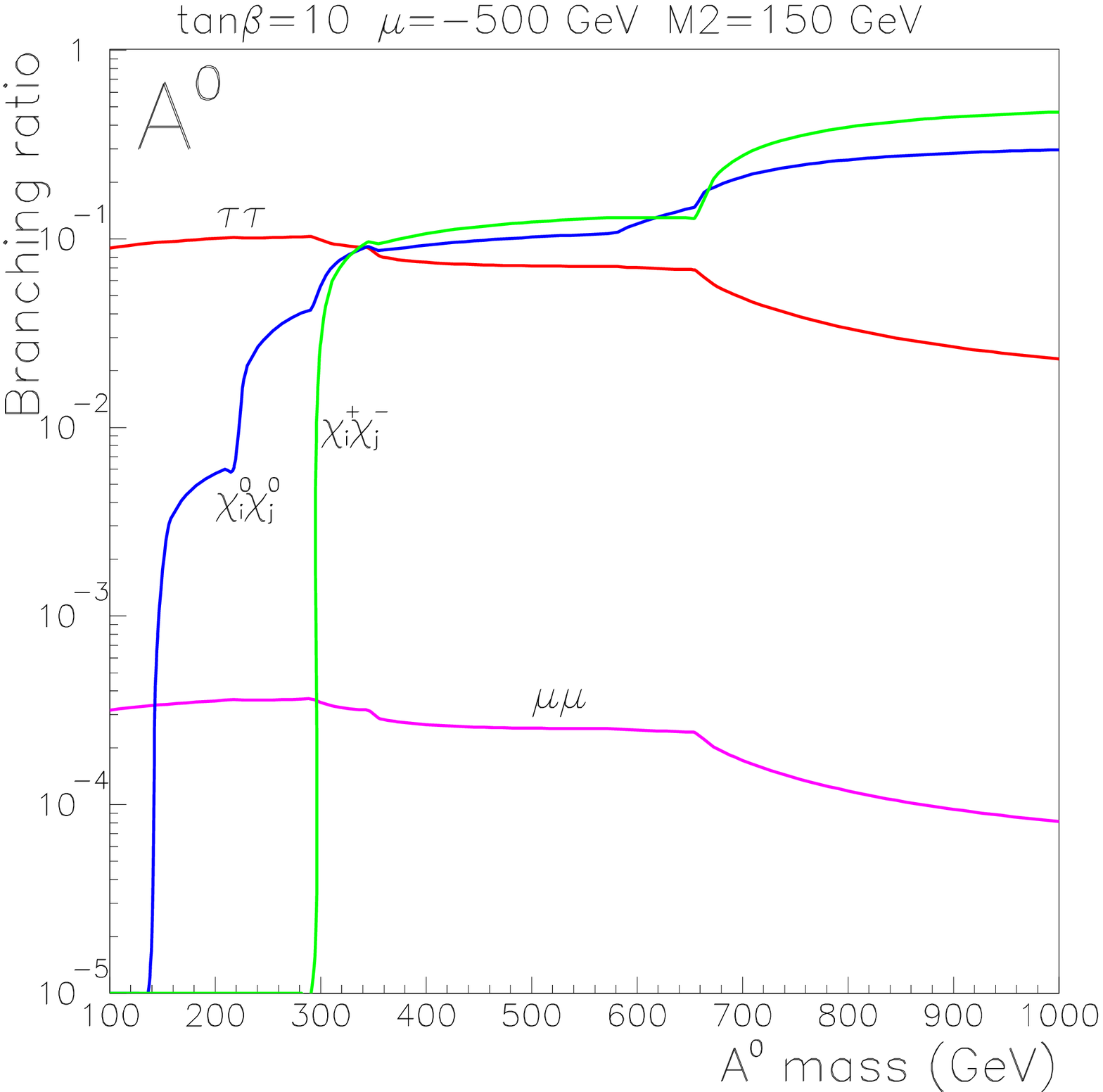,height=60mm}
\caption{Example of the branching ratios of $A^0$ into $\tau \tau$, $\mu \mu$ and sparticles.
\label{fig:1}}
\end{center}
\end{minipage}
\begin{minipage}[l]{10mm}
\hspace{10mm}
\end{minipage}
\begin{minipage}[l]{65mm}
\begin{center}
\epsfig{file=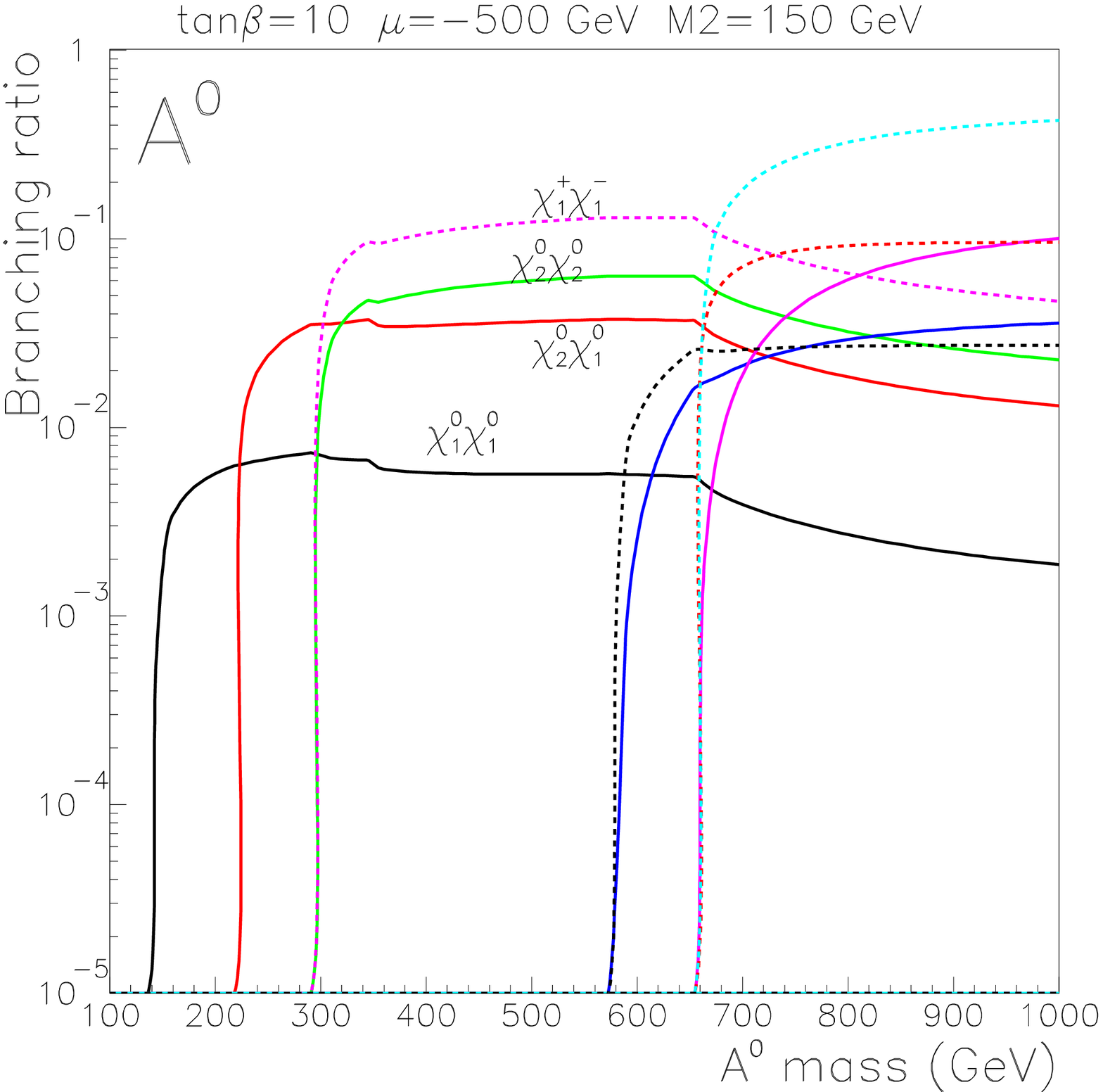,height=60mm}
\caption{Detailed branching ratios of $A^0$ into neturalinos and charginos.
\label{fig:2}}
\end{center}
\end{minipage}
\end{figure}

\begin{figure}
\begin{minipage}[l]{65mm}
\begin{center}
\epsfig{file=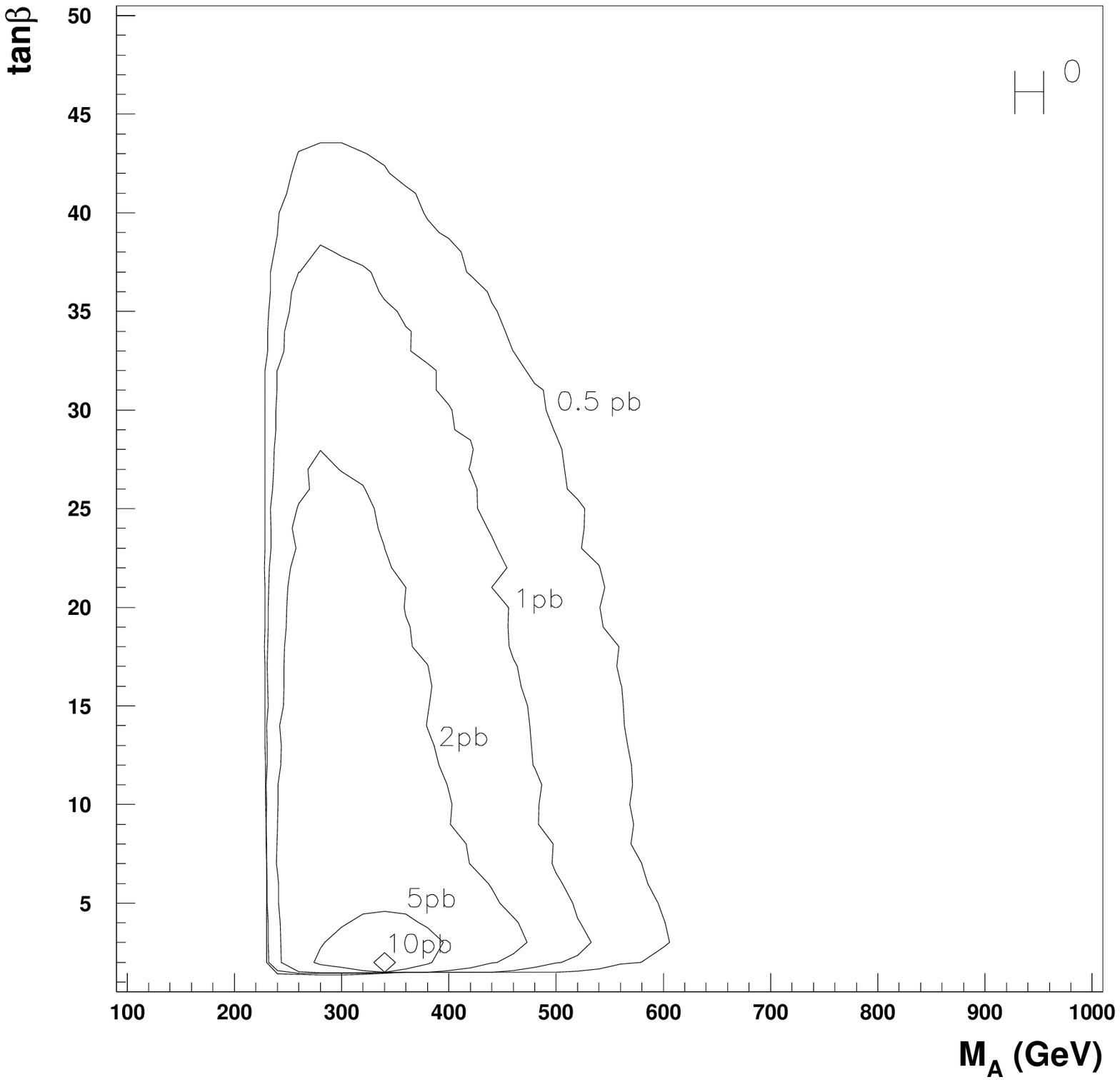,height=65mm}
\caption{$\sigma \times$ BR contours for $H^0$ in the $m_A-\tan\beta$ plane. MSSM parameter values as described in the text.
\label{fig:3}}
\end{center}
\end{minipage}
\begin{minipage}[l]{10mm}
\hspace{10mm}
\end{minipage}
\begin{minipage}[l]{65mm}
\begin{center}
\epsfig{file=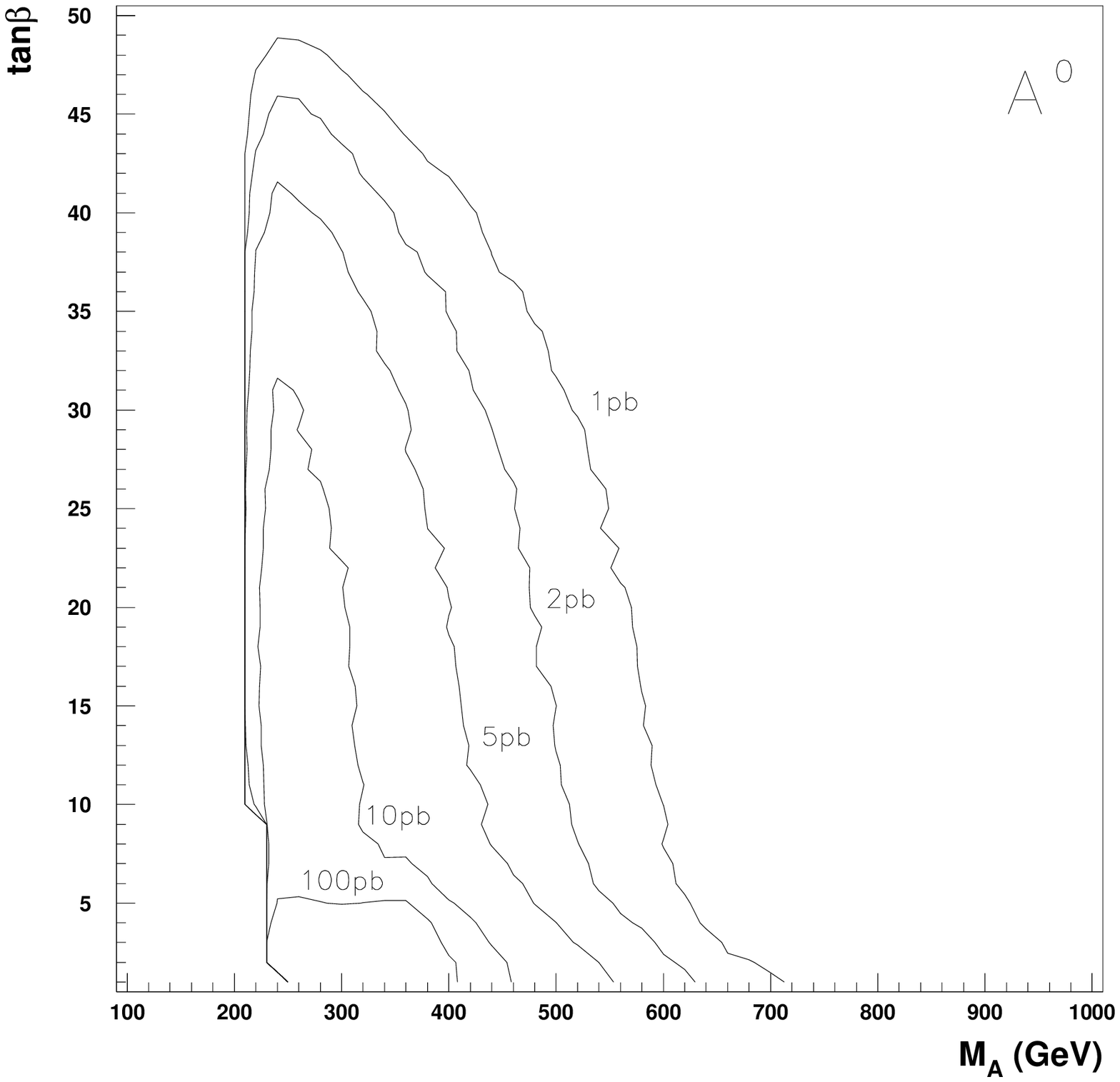,height=65mm}
\caption{$\sigma \times$ BR contours for $A^0$ in the $m_A-\tan\beta$ plane. MSSM parameter values as described in the text.
\label{fig:4}}
\end{center}
\end{minipage}
\end{figure}

The supersymmetric decay modes of the heavy Higgses were first studied in the
MSSM framework using the
HDECAY \cite{hdec} package. Fig. \ref{fig:1} shows the branching ratios of $A^0$ into the $\tau \tau$, the $\mu \mu$ and the 
sparticle modes, for the chosen set of MSSM parameters. For $m_A$ $\lesssim$ 500 GeV, 
the decay probability of the heavy Higgs into neutralinos and charginos can be as high as 20\%. The other
sparticle decay modes seem to be neglegible.
Looking more in detail to the chargino/neutralino modes (fig. \ref{fig:2}), one sees that the 
$\chi_1^+\chi_1^-$ decay mode gives the highest branching ratio (for $m_A$
$\lesssim$ 500 GeV), while the second best sparticle mode is $\chi^0_2 \chi^0_2$.
In view of the large backgrounds at the LHC, the leptonic decay modes 
of the neutralinos/charginos seem most preferable: $\chi^0_2 \ra l^+ l^- \chi^0_1$ 
and $\chi^{\pm}_1 \ra l^{\pm} \nu \chi^0_1$.
The amplitude of the $\chi^0_2 \ra l^+ l^- \chi^0_1$ decay strongly depends on the values of 
$M_1$, $M_2$, $\tan\beta$, $\mu$ and $m_{\tilde{l}}$. This is due to the fact that 
for the decay to leptons, two diagrams contribute: the $Z^0$ exchange and the virtual 
slepton exchange. Depending on the value of $\mu$, it may be more favourable to have 
light or heavy sleptons, because the diagrams can interfere positively as well as negatively.
However, in general the branching ratio into electrons and muons will decrease with $\tan\beta$ due to the enhanced 
coupling with taus. \\
In order to have a strong experimental signature allowing to suppress the backgrounds, we will consider the channel
\begin{equation}
A^0, H^0 \ra \chi^0_2 \chi^0_2 \ra 4l^{\pm} + X
\end{equation}
with four {\em isolated} leptons ($e$, $\mu$) in the final state.
\\
The $A^0, H^0 \rightarrow  \chi^0_2 \chi^0_2 \rightarrow 4 l^{\pm}$ cross section was
first scanned 
in the $m_A$ - $\tan\beta$ plane for different values of $M_2$, $\mu$ and $m_{\tilde{l}}$. In 
figs. \ref{fig:3} and \ref{fig:4}, the plot 
of $\sigma \times$ BR for $M_1$ = 60 GeV, $M_2$ = 110 GeV, $\mu$ = -500 GeV and $m_{\tilde{l}}$ 
= 250 GeV is shown for the $H^0$ and $A^0$. Values of $\tan\beta$ $\lesssim$ 30
and $m_A$ $\lesssim$ 400
GeV are favoured. One also notices that the pseudoscalar Higgs gives much 
higher cross sections than the scalar one, due to its stronger coupling to $b\bar{b}$
(for $\tan\beta$ $\gtrsim$ 5, the associated production mechanism $gg, q\bar{q} \ra A^0, H^0
b\bar{b}$ dominates over the gluon fusion $gg \ra A^0, H^0$).\\
We will first choose a favourable point in parameter space and study whether we
can observe this $4l^{\pm}$ signal above the expected backgrounds.
\subsection{Event generation}
The signal $A^0, H^0 \ra \chi^0_2 \chi^0_2 \ra 4l^{\pm}$ was generated with SPYTHIA
\cite{spythia}.
The signature contains two pairs of leptons with opposite sign and same 
flavour, in addition to a substantial amount of missing energy due to the 
escaping lightest neutralinos.\\
Possible backgrounds that can mimick this signature are
$ZZ$, $ZW$, $Zb\bar{b}$, $Zc\bar{c}$, $Wt\bar{b}$ and $t\bar{t}$ final states in the
Standard Model and $\tilde{q}/\tilde{g}$, $\tilde{l}\tilde{l}$, 
$\tilde{\nu}\tilde{\nu}$, $\tilde{q}\tilde{\chi}$,
$\tilde{\chi}\tilde{\chi}$ production in the MSSM extension.\\
The background events are generated with PYTHIA 6.136 \cite{pythia}.
The CMS detector response was simulated using the fast simulation package 
CMSJET \cite{cmsjet}.

\subsection{Event selection}

We can first discriminate between the signal and the background by 
making the following basic requirements:
\begin{itemize}
\item there must be four {\em isolated} leptons ($e$/$\mu$) in the final state, with a 
transverse momentum higher than 10 GeV and with $\eta$ $<$ 2.4. We demand a tight isolation  
of the leptons in the tracker (no charged particle with $P_T$ $>$ 1.5 GeV in the cone of 0.3 rad around each
 lepton track) as well as in the electromagnetic calorimeter (the sum of the transverse energy 
 in the crystal towers between 0.05 and 0.3 rad around the track has to be smaller than 3 GeV)
\item we only consider lepton pairs with dilepton effective mass outside the
range $m_Z \pm$ 10 GeV (Z veto).
\end{itemize}
These conditions effectively suppress the SM backround processes. 
The explicit Z veto is used in order to maximally suppress the $ZZ$ and $WZ$
backgrounds,
while the $t\bar{t}$ background practically disappears if the lepton isolation
criterion is tight enough.  \\
Further optimisation of the signal to background ratio can be obtained by requiring that:
\begin{itemize} 
\item $P_T$ hardest lepton $<$ 80 GeV
\item 20 Gev $<$ $E_T^{miss}$ $<$ 130 GeV
\item $E_T$ hardest jet $<$ 100 GeV
\item if necessary (light squark/gluino): only 0,1,2 jets
\end{itemize}
The only important SM background process still surviving these requirements is ZZ production, 
mainly when one of the $Z$ bosons decays into taus. 
Most of the background events surviving the selection are supersymmetric
in nature: sneutrino and neutralino direct pair production. 

\subsection{Case studies}
We focus on two points in the MSSM parameter space. The first one is the point where
the signal cross section is optimal. In the second point, the case of heavier neutralinos 
and a positive value for $\mu$ is considered.
\subsubsection{Case 1:}
\begin{figure}
\begin{minipage}[l]{65mm}
\begin{center}
\epsfig{file=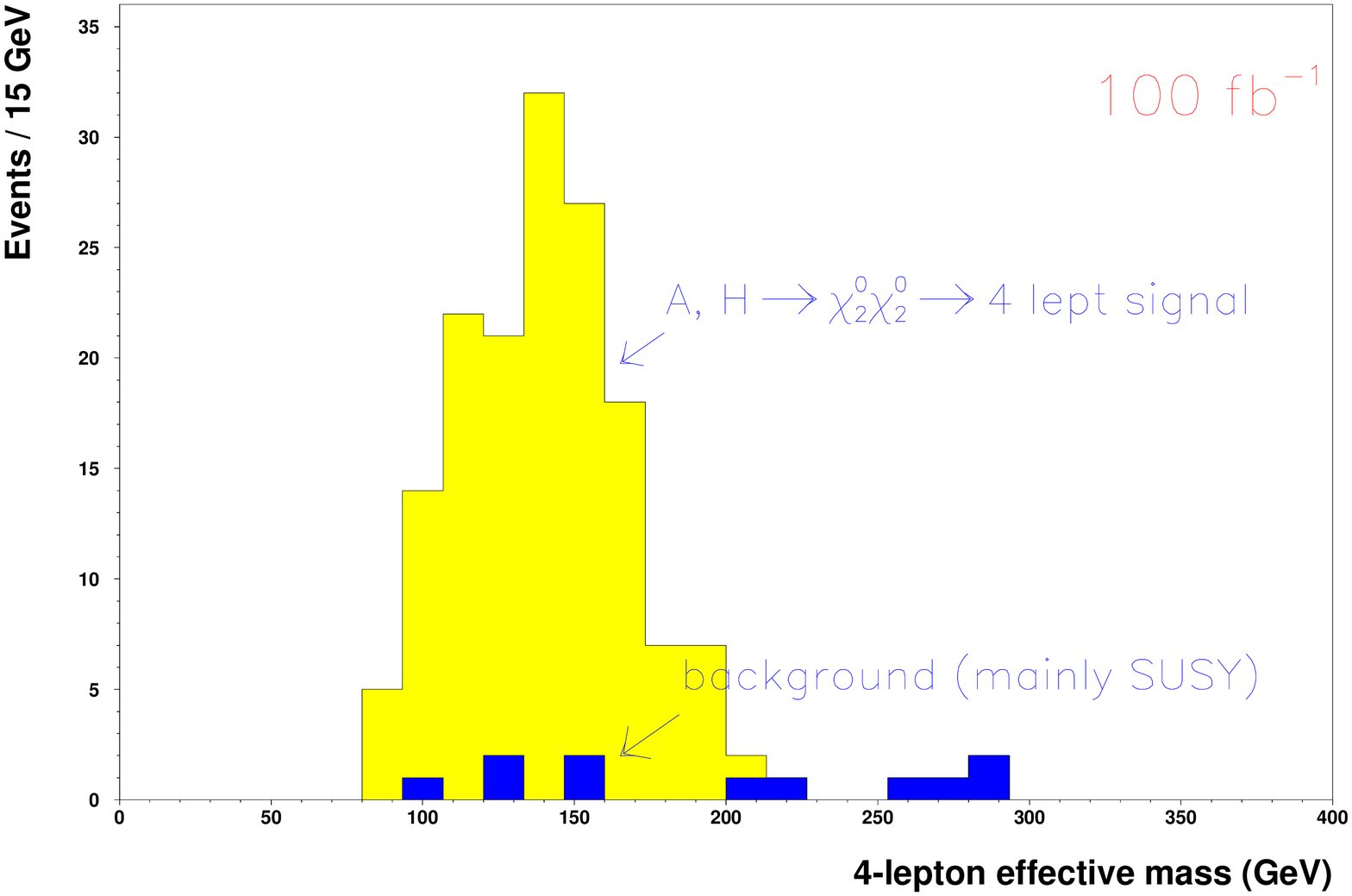,height=65mm, width=65mm}
\caption{Signal over background for {\em case 1} in the 4-lepton effective mass plot. MSSM parameters as described in the text.
\label{fig:5}}
\end{center}
\end{minipage}
\begin{minipage}[l]{10mm}
\hspace{10mm}
\end{minipage}
\begin{minipage}[l]{65mm}
\begin{center}
\epsfig{file=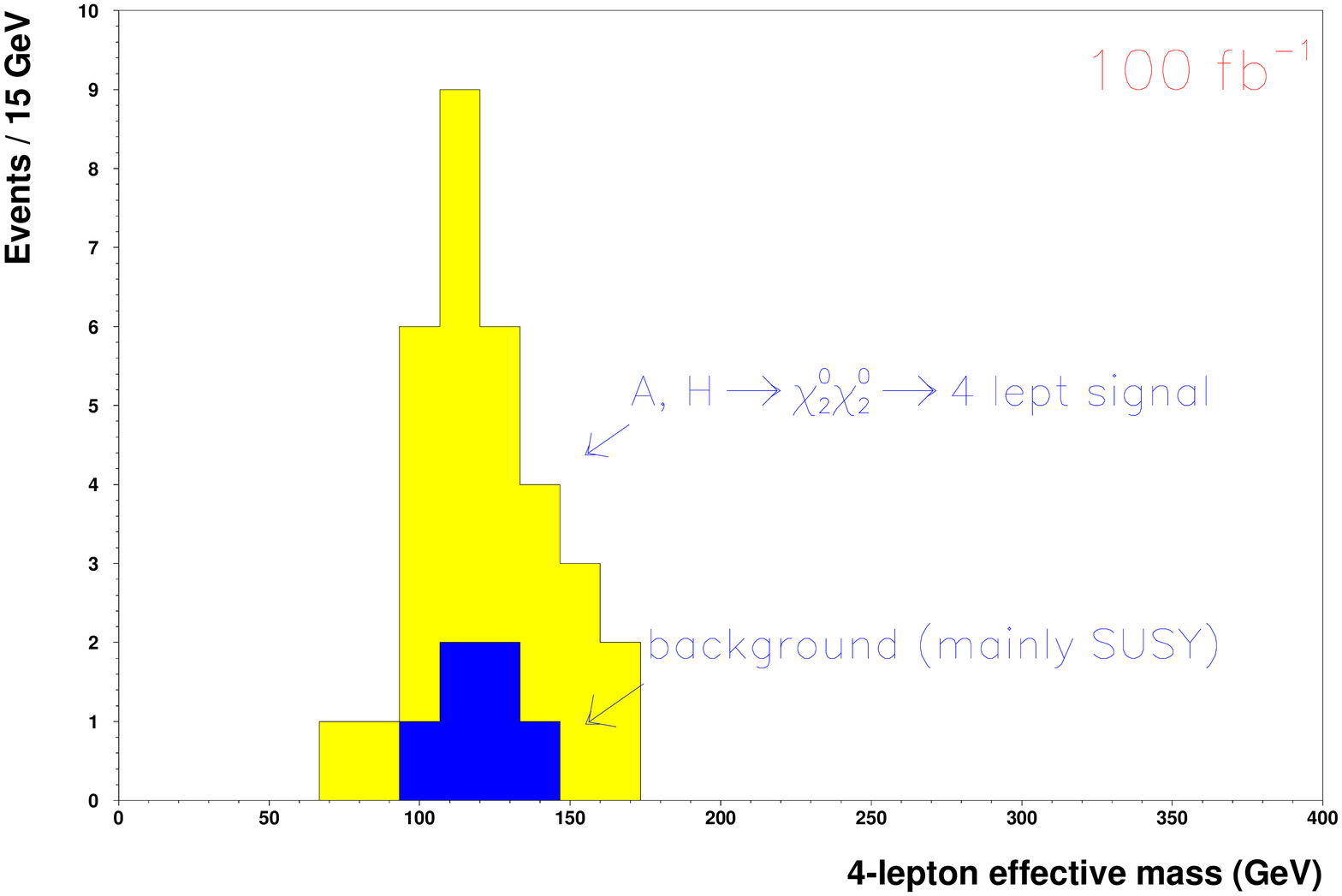,height=65mm, width=65mm}
\caption{Signal over background for {\em case 2} in the 4-lepton effective mass plot. MSSM parameters as described in the text.
\label{fig:6}}
\end{center}
\end{minipage}
\end{figure}

\begin{figure}
\begin{center}
\epsfig{file=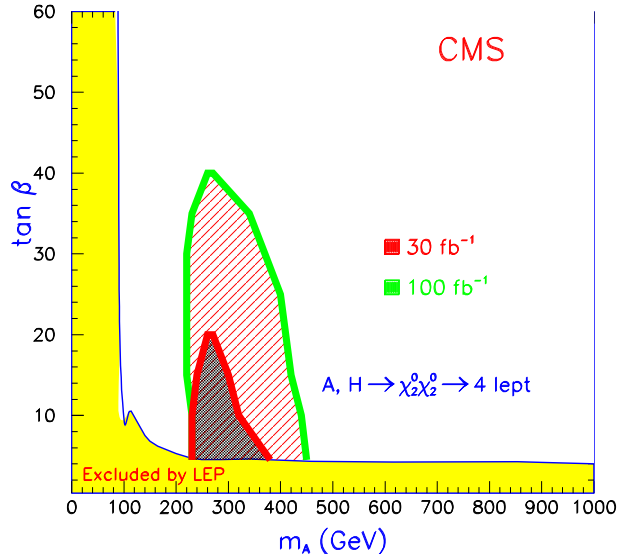,height=80mm}
\caption{5$\sigma$ discovery contours for {\em case 1} at 30 and 100
fb$^{-1}$.                                         
MSSM parameters: $M_2$ = 120 GeV, $M_1$ = 60 GeV, $\mu$ = -500 GeV, $m_{\tilde{l}}$ = 250 GeV, $M_{\tilde{q}, \tilde{g}}$ = 1000 GeV.
\label{fig:7}}
\end{center}
\end{figure}

The following values for the MSSM parameters were chosen: $M_2$ = 120 GeV, $M_1$ = 60
GeV,\\
$\mu$ = -500 GeV, $m_{\tilde{l}}$ = 250 GeV, $M_{\tilde{q}, \tilde{g}}$ = 1000 GeV.
In this case we can obtain a clear signal over the background, like in fig.
\ref{fig:5} for $m_A$ = 320 GeV and $\tan\beta$ = 5 (best case).\\ 
In fig. \ref{fig:7}, the region in the $m_A-\tan\beta$ plane is plotted where a $5\sigma$-discovery 
could be made for an integrated luminosity of 30 and 100 $fb^{-1}$, provided we understand
the nature of both SM and SUSY backgrounds in the final state topology. 
The discovery region starts where the $\chi^0_2\chi^0_2$ decay becomes kinematically 
accessible, $m_A \, \ge \, 2m_{\chi_2^0}$ = 225 GeV;  
the upper reach in $m_A$ is mainly determined by the $A/H$ production cross section, which drops
with $m_A$ as a power law. The reach in $\tan \beta$ is determined by the branching ratio of $\chi_2^0
\rightarrow l^+ l^- \chi_1^0$.
At 30 $fb^{-1}$, the discovery region reaches $m_A \sim $ 350 Gev and $\tan\beta \sim$ 20.
For 100 $fb^{-1}$, we see that values of $\tan\beta$ $\sim$ 40 and masses up to $m_A \sim$ 450 GeV are 
accessible.
This area is - at least partly - covering the difficult region of MSSM parameter space 
that is not easily accessible for SM decays of SUSY Higgses
- except for the $h \ra b\bar{b}$ mode.
\subsubsection{Case 2:}
The following MSSM parameter choices were made: $M_2$ = 180 GeV, $M_1$ = 100
GeV,\\
$\mu$ = +500 GeV, $m_{\tilde{l}}$ = 250 GeV, $M_{\tilde{q}, \tilde{g}}$ = 1000 GeV.
Here, the mass of the next-to-lighest neutralino will be larger, so the $\chi^0_2 \chi^0_2$ 
mode will only start being accessible around $m_A$ $\sim$ 350 GeV. 
In fig. \ref{fig:6} the expectations are shown for $m_A$ = 380 GeV and
$\tan\beta$ = 10. The excess of the signal over the background is less
pronounced than in {\em case 1}, but still very clearly visible.\\
\\
Due to the $\chi^0_2 \ra l^+ l^- \chi^0_1$ decay, there will be a clear 
"kinematical edge" at the mass difference $M_2$ - $M_1$ in the dilepton effective mass 
spectrum for the signal events. Even a double kinematic edge 
is visible in the di-electron versus the di-muon effective mass plot.
Therefore, an observation of the signal would also allow us to determine extra parameters 
in the theory.
\section{Conclusions and Outlook}
The channel $H^0, A^0 \rightarrow \chi^0_2 \chi^0_2 \rightarrow
4 \, l^{\pm} \,$, with four isolated leptons in the final state, may be observed in the low and intermediate $\tan\beta$ 
region of the MSSM parameter space, if neutralinos and sleptons are light enough.
This region is largely complementary to the reach of the $H^0,A^0 \rightarrow \tau\tau$ 
channel. The discovery potential will depend on other MSSM parameters like
$m_A$, $\tan\beta$, $M_1$, $M_2$, $\mu$ and $m_{\tilde{l}}$.
We are currently mapping the reach in $m_A$-$\tan\beta$ parameter space varying 
these parameters.
As a next step, one might ask whether it would be possible to observe the supersymmetric 
decay of the charged Higgs $H^+ \rightarrow \chi^0_2 \, \chi^{\pm}_{1,2}$ at the LHC.
In order to suppress the background, a final state with three isolated leptons seems 
preferable. Since the charged Higgs is mainly produced via $gb \rightarrow tH^+$, 
we can also try to reconstruct the associated top quark for additional background 
suppression. Investigations of this topic are under way.

\end{document}